\documentclass[]{spie}  
\usepackage[]{graphicx}
\usepackage{aas_macros}

\title{A Dispersed Heterodyne Design for the Planet Formation Imager\footnote{\small~Copyright 2014 Society of Photo-Optical Instrumentation Engineers. One print or electronic copy may be made for personal use only. Systematic reproduction and distribution, duplication of any material in this paper for a fee or for commercial purposes, or modification of the content of the paper are prohibited. DOI abstract: http://dx.doi.org/10.1117/12.2057355}} 

\author{Michael J. Ireland\supit{a} and John D. Monnier\supit{b}
\skiplinehalf
\supit{a}Research School of Astronomy \& Astrophysics, Australian National University, Canberra ACT 2611, Australia; \\
\supit{b}Department of Astronomy, University of Michigan, 941 Dennison Building, Ann Arbor, MI 48109, USA
}

\authorinfo{Further author information: M.J.I.: E-mail: michael.ireland@anu.edu.au}

\begin{document} 
  \maketitle 

\begin{abstract}
The Planet Formation Imager (PFI) is a future world facility that will image the process of planetary formation. It will have an angular resolution and sensitivity sufficient to resolve sub-Hill sphere structures around newly formed giant planets orbiting solar-type stars in nearby star formation regions. We present one concept for this design consisting of twenty-seven or more 4m telescopes with kilometric baselines feeding a mid-infrared spectrograph where starlight is mixed with a frequency-comb laser. Fringe tracking will be undertaken in H-band using a fiber-fed direct detection interferometer, meaning that all beam transport is done by communications band fibers. Although heterodyne interferometry typically has lower signal-to-noise than direct detection interferometry, it has an advantage for imaging fields of view with many resolution elements, because the signal in direct detection has to be split many ways while the signal in heterodyne interferometry can be amplified prior to combining every baseline pair.
 We compare the performance and cost envelope of this design to a comparable direct-detection design.
\end{abstract}



\section{INTRODUCTION}
\label{sec:intro}  

The Planet Formation Imager (PFI) is a future world facility that will image the process of planetary formation, especially the formation of giant planets. It will complement the Atacama Large Millimetre Array (ALMA) which will image the influence of planetary formation on disks, instruments like SPHERE and GPI, which will image mature planetary systems, and long-term radial velocity monitoring programs, which are now beginning to collect statistics on Jupiter analogs\cite{Wittenmyer14}. PFI will have an angular resolution and sensitivity sufficient to resolve structures smaller than the sphere of gravitational influence of a newly formed giant planet - its {\em Hill sphere} defined by: 

\begin{equation}
r_H = a \sqrt[3]{\frac{m_p}{M_s}},
\end{equation}

where $a$ is the semi-major axis of the planetary orbit, $m_p$ is the planet mass and $M_s$ is the star mass. This resolution requirement for a Jupiter-mass planet at a 1\,AU separation is approximately 0.05 AU. In order to truly image the range of separations most likely to form planets, a 20 AU field of view is required\cite{Kraus12}, and the instrument would have to be sensitive enough to carry out a comprehensive survey of solar-type stars in nearby star forming regions at ~140\,pc (e.g. Upper Scorpius and Taurus). For the purposes of this paper, we will assume that these science requirements apply as a minimum  to the astronomical N-band (10--13\,$\mu$m) as justified below, and are taking the following as the instrumental requirements:

\begin{itemize}
\item Fringe tracking for mid-infrared coherencing as faint as H=9.0, N=7.5 (a young solar-type star, e.g. LkCa~15, in Taurus or Upper Scorpius).
\item 1-$\sigma$ sensitivities of N=15 in 10 hours of integration, including any ``speckle-noise'' or cross-talk, at a working angle of 7 milli-arcsec or further from an N=7.5 central star.
\item An angular resolution of 0.35\,milli-arcsec or better (0.05\,AU at 140\,pc).
\item An imaging field of view of at least 140\,milli-arcsec radius.
\end{itemize}

The details of these requirements will be debated over the coming months and years by the science working group of the Planet Formation Imager collaboration, but these are a at least one set of requirement for a fully-functional instrument. We note that the adaptive optics imaging work of e.g. Follette et al\cite{Follette13} for the SR~21 transitional disk system shows that, at least at moderately low spatial resolution, scattering by small grains can give a relatively large signal from complex structures not necessarily associated with a disk or planets. Thermal radiation from a newly-formed planet and its environment should overwhelm scattered light from low-mass optically-thin dust structures, and it is the thermal radiation that we will focus on in this paper. 

\begin{figure}
\includegraphics[width=\textwidth]{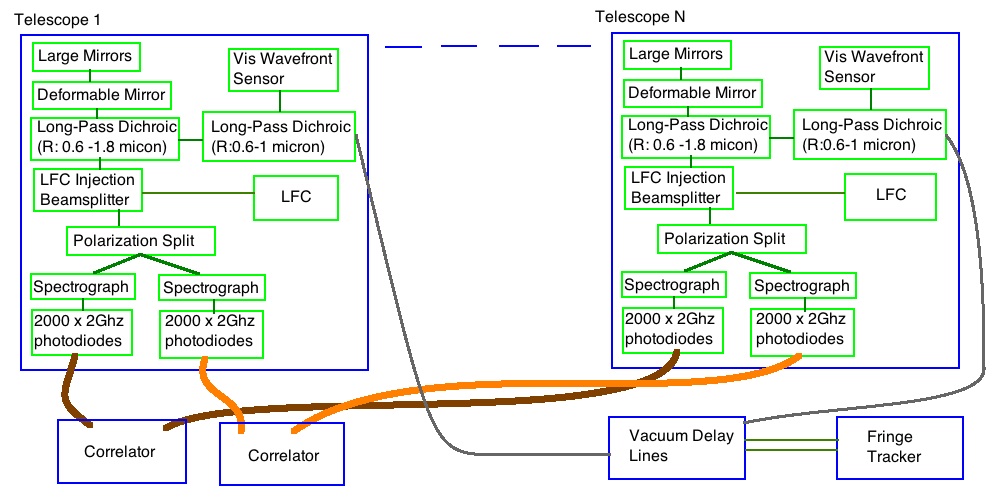}
\caption{A system level diagram of the heterodyne PFI concept. For each of N telescopes, the wavelengths are split in 3: 10--13\,$\mu$m radiation for heterodyne interferometry, 
1.1--1.8\,$\mu$m radiation for fringe tracking, and 0.6--1.0\,$\mu$m radiation for adaptive optics wavefront sensing. The 1.1--1.8\,$\mu$m radiation is transported by fiber to long
vacuum delay lines and a conventional fringe tracker. The 10--13\,$\mu$m radiation is mixed with a laser frequency comb, split in polarisation and sent through a high-resolution 
spectrograph. The 4\,THz output bandwidth of each pair of telescopes, for each polarisation is correlated in real time, using the 1.1--1.8\,$\mu$m fringe tracker for post-processing 
coherencing.}
\label{figSchematic}
\end{figure}

The luminosity output by a forming proto-Jupiter is thought to be of order $2 \times 10^{-4}$\,L$_{\odot}$, over a timescale of a few $\times 10^5$ years\cite{Lissauer09}. The planetary temperature is of order 1500\,K with a background temperature in the disk of order 150\,K, a stellar temperature of order 4000\,K and a stellar luminosity of order 1\,$L_\odot$. Naeively, this would correspond to a required contrast of $4 \times 10^{-4}$ at the blackbody peak at $\sim$2\,$\mu$m (8.4 magnitudes) and $3 \times 10^{-3}$ at 11.5\,$\mu$m (6.4 magnitudes), however this does not take into account re-processing of radiation by the circumplanetary disk and environment, which makes the longer wavelengths more suitable. If the complete $\sim2 \times 10^{-4}$\,L$_{\odot}$ is re-processed by a 500\,K disk, then the disk/star ratio at 11.5\,$\mu$m could be as much as 0.05 (3 magnitudes). Conversely, this reduction in effective temperature of the planet to 500\,K by circumplanetary material (still well within the Hill sphere) would be enough to render 2\,$\mu$m useless, although we concede that alternative designs focusing on the 3--4\,$\mu$m region may also be suitable, and would be less influenced by the background radiation from the protoplanetary disk. 

The resolution requirement of one Hill sphere radius is not just to enable the separation of signals from neighbouring planets and to maximise the ability to detect bright point-source like emission from the planet. It is also essential to separate the planetary emission from background disk thermal radiation. If, for example, a planet is forming in a disk gap that is only as wide as its Hill sphere, then the planetary signal would rapidly become dwarfed by the emission from the disk at a few Hill radii. 

In this paper, we argue that a heterodyne design is competitive to direct detection designs for detecting this thermal radiation and should be considered as one of the options for PFI. This design has a low-order adaptive optics system creating a collimated beam which is mixed with a laser frequency comb. This mixed signal is then fed into a spectrograph disperses the mixed signal onto an linear array of individual detectors, ideally one for each polarisation. We begin by describing the overall interferometer architecture and sensitivity, then move into plausible designs for the individual components. A schematic system diagram is given in Figure~\ref{figSchematic}.

\section{ARRAY FORMAT}

\begin{figure}
\includegraphics[width=0.5\textwidth]{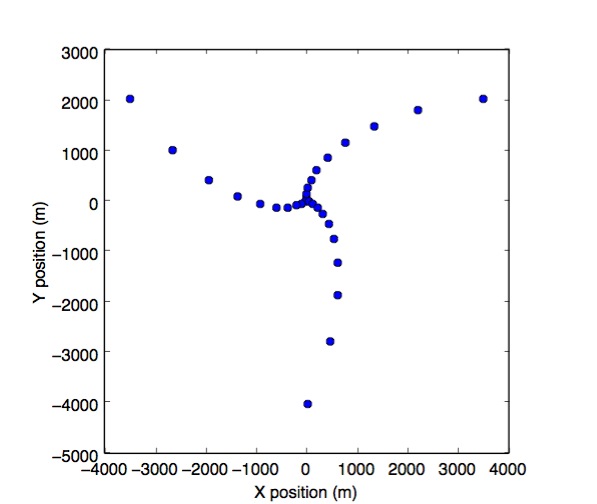}
\caption{An example 30-telescope array configuration, with maximum baseline 7\,km, minimum baseline 82\,m and excellent imaging and baseline-bootstrapping properties.}
\label{figFormat}
\end{figure}

At a central N-band wavelength of 11.5\,$\mu$m, the angular resolution requirement (taken to be $\lambda/B$) corresponds to maximum baseline of 7\,km. Allowing for some super-resolution might enable baselines to be reduced to 3\,km, but not without a reduction in sensitivity due to possible contamination by the disk emission as discussed above. The imaging requirement for PFI is essential because complex structures in the disk could mask planetary signals if ambiguous modelling of interferometric data are needed. The minimum number of baselines is roughly the number of resolution elements across the final image - 400 in our case. This drives PFI to consider a large number of telescopes - of order 30 or more, noting that the number of baselines $N_B = N_T (N_T-1)/2$ with $N_T$ the number of telescopes. This number of telescopes is inadequate for snapshot imaging, and requires earth rotation synthesis to get the full $\sim$400 x 400 resolution element image. Snapshot imaging fidelity is arguably not required, as the smallest timescale required to be resolved is $\sim$ half the rotation period of the proto-Jupiter, or a few hours. Given the need for ground-based observations sensitive to atmospheric conditions to avoid high air-masses, it is also important that earth rotation synthesis can be achieved in only a few hours of observing and not a full night. Taken together, these requirements mean that a 3-arm spiral (e.g. Figure~\ref{figFormat}) or a Y-shaped array with a format similar to the VLA is close to optimal.

The aperture size of the individual telescope elements also can not be so large that the $\sim$20\,AU radius field of view is outside the primary beam. For a 11.5$\mu$m wavelength, a requirement of 20\,AU/140\,pc $< \lambda/D$ means that the telescope diameter of individual array elements should not be larger than 8\,m. 

\section{SENSITIVITY}

Perhaps the most important requirement for sensitivity is the telescope diameter. For an array limited by the number of telescope elements (e.g. if the beam-combiners and delay-line architecture are most important in the total cost), the appropriate question is : `` what is the minimum telescope size required to achieve the sensitivity goals?". For an array where the cost is driven by the telescope cost, we note that more small telescopes are cheaper\cite{vanBelle04}, so the appropriate question is: ``what is the number of telescopes of fixed size required to achieve the sensivity goals?". We will consider both these questions - the minimum telescope diameter if there are 30 telescopes in the array, and the number of telescopes in the array if they are all 2\,m telescopes.

The sensitivity of Heterodyne detection is given by\cite{Hale97}:

\begin{equation}
 \left( \frac{S}{N} \right )_h = \frac{P_\nu}{h\nu} \sqrt{2 \Delta \nu t},
\end{equation}

where $P_\nu$ is the received power per polarisation per unit bandwidth (Hz). Note that the thermal background does not appear in this equation, because the quantum noise for heterodyne detection is equivalent to a thermal background of $T=h\nu/k_B$, or 1250\,K at 11.5\,$\mu$m. 

A simple thought experiment where a large aperture is split into many small ones, and the voltages added prior to forming an autocorrelation, makes it easy to see that this formula applies to an interferometer or to single-dish heterodyne detection, so long as al the amplitude, phase and auto-correlation information is available. In the case of unavailable (or useless for imaging) auto-correlations and antenna phases, this signal-to-noise is reduced by a factor of $(N_T-1)/N_T$. Taking the 11.5\,$\mu$m flux of Vega as 17\,Jy per polarisation, we then arrive at the overall signal-to-noise per polarisation at 11.5\,$\mu$m:

\begin{equation}
 \left( \frac{S}{N} \right )_h = 9.8 \times 10^{-6} {\rm m}^{-2} A_T \eta (N_T-1) \sqrt{2 \Delta \nu t}~10^{-0.4 m_N},
\end{equation}

where the total collecting area per telescope is $A_T$, the target Vega magnitude is $m_N$ and the system efficiency is $\eta$. For a system bandwidth $\Delta \nu$=4\,THz and $\eta=0.35$ with a dual-polarisation design (or $\eta=0.5$ with a single-polarisation design), the requirement for (S/N)=1 for $m_N = 15$ in 10 hours gives a minimum telescope diameter of 4\,m for $N_T$=30. With the same bandwidth and system efficiency, but 2\,m telescopes, the minimum number of telescopes is 136. If the efficiency $\eta$ is halved, then the required number of telescopes doubles. This clearly shows that the heterodyne efficiency must be maximised.

\section{SPECTROGRAPH DESIGN}

Starlight from a low-order adaptive optics system is first mixed with the Laser Frequency Comb (LFC) with 3\,GHz line spacing via a beamsplitter of low (e.g. 5\%) reflectivity, that reflects the 5\% of wasted starlight toward a cold stop. Given that gratings are only efficient in one polarisation, the most efficient time to split the polarisations is immediately after mixing with the LFC. We note that if e.g. the LFC is circularly polarised, it will effectively mix with both outputs of a linear polarising beamsplitter.

\begin{figure}
\includegraphics[width=0.7\textwidth]{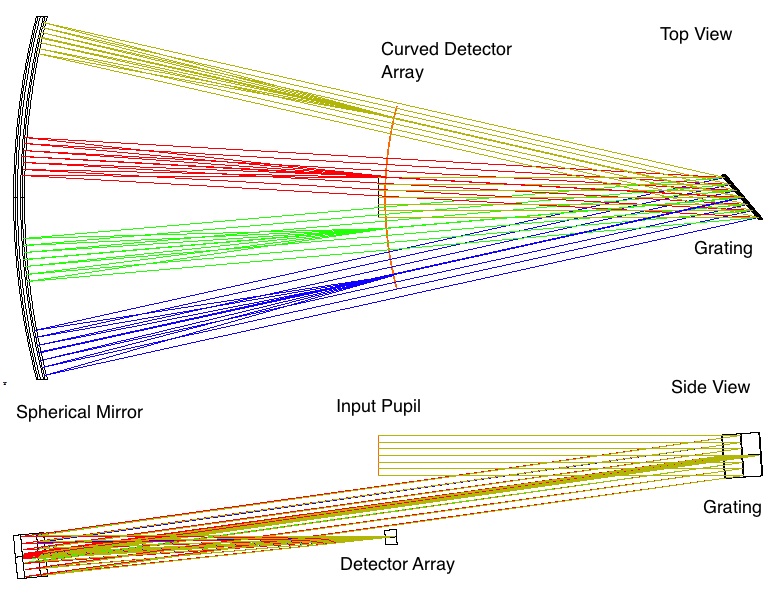}
\caption{An example spectrograph configuration that would give $\sim$0.4mm physical separation between laser lines and would be able to have the appropriate resolving power of R$\sim$17000 in order to separate the
comb lines of 3\,GHz spacing. The input pupil in this case is 110\,mm, the focal-length of the spherical mirror 1\,m and the grating Littrow angle 40 degrees. Pupils as large as 200\,mm could be accommodated in this arrangement with smaller grating angles, but would require more space.}
\label{figSpectrograph}
\end{figure}

This mixed radiation is then dispersed using a spectrograph that has a grating used at 40 degree angle of incidence with a 110mm pupil, with each comb line landing on a separate single cooled HgCdTe detector with a 2 GHz bandwidth. This is roughly the maximum achievable bandwidth for a HgCdTe photodiode\cite{Rogalski05}. Although not simple mechanically, a large number of these photodiodes (2000 in our case) could be placed next to each other, injected by individual micro lenses, and a separate 2\,GHz output extracted from each photodiode. An example configuration is shown in Figure~\ref{figSpectrograph}.

\begin{figure}
\includegraphics[width=0.6\textwidth]{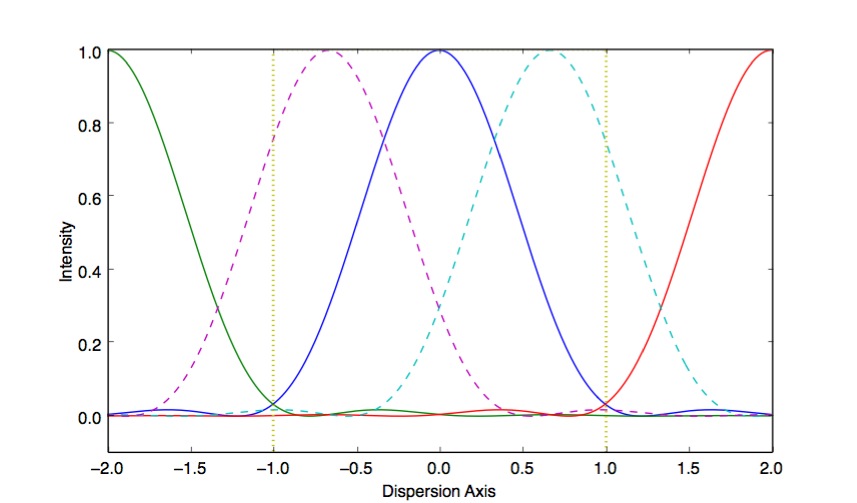}
\caption{The dispersed image-plane, showing the comb lines (solid lines), starlight at the +/- 1\,GHz bandpass edges (dashed lines) and the edges of one resolution element (dotted lines).}
\label{figOverlap}
\end{figure}

Unfortunately, the starlight landing on each individual detector element is dispersed, so that only the wavelengths that exactly match the comb lines will have a perfect spatial overlap. The example spectrograph here would have R$\sim$17000, which is enough to have the comb-lines separated by $2\lambda/D_{\rm pupil}$, with the starlight 1\,GHz from each comb line having a reduced overlap integral by a factor of 0.65. There are also small effects of cross-talk (see Figure~\ref{figOverlap}).

\section{LASER FREQUENCY COMB}

Laser frequency combs, in development for more then 15 years\cite{Holzwarth00} are now available commercially in the visible and near-infrared wavelength ranges. In order to produce a mid-infrared laser frequency comb, one solution is to mix a 1.55\,$\mu$m now commercially available comb (e.g. the FC1500-250-WG from Menlo Systems) filtered by a Fabry-Perot\cite{Steinmetz09}, with a stabilised tuneable ~1400nm diode laser. Difference frequency generation is achievable using non-linear materials such as GaSe, producing comb lines with a target 3 GHz separation throughout the 10-13\,$\mu$m bandpass. We estimate that a conversion efficiency of $\sim$0.1\% is achievable, which could produce power after of order 1\,$\mu$W per comb line after mixing. Over a 2\,GHz bandwidth, the laser power is not a significant source of noise as long as many more than 2$\times 10^9$ photons/s are detected by each HgCdTe photodiode - equivalent to a laser power of $>$30\,pW, a requirement which is relatively easy to meet. The more relevant requirement for the LFC and detector system is that the Noise Equivalent Power (NEP) due the quantum noise from the laser fluctuations in each comb mode far exceeds the NEP of the detector itself. This is certainly achieved by 77\,K photodiodes of $<$0.1\,mm size and NEP values less than $10^{-13}$\,W\,Hz$^{1/2}$ and comb powers of order 1\,$\mu$W per comb line, but it could prove to be a difficult requirement to meet with a full system model and high-bandwidth photodiodes.

\section{CORRELATOR REQUIREMENTS}

A key difference between this correlator and radio correlators is that there may be no need for many channels within each 2\,GHz bandpass, as the $\sim$30\,km\,s$^{-1}$ resolving power of each 2\,GHz channel is at least enough to separate strong lines form the continuum, and to measure spectral features due to dust. With a 30 telescope design, the 4\,THz full bandwidth of a heterodyne PFI requires $\sim$350 times more processing power than ALMA, i.e. correlating few-bit samples at 2.8\,PHz. Assuming that a PHz for few-bit samples equivalent to a Pflop, this processing requirement is already less than the 10 most powerful supercomputers today. The power requirement (approaching the dominant cost for super-computers) is approximately 2\,MW, which is smaller than typical operating costs of a typical single large telescope today. For this reason, especially once typical computer speed and power requirement scaling laws are taken into account, we do not expect the correlator to be a significant cost component to a heterodyne PFI for a construction date $\sim$10 or more years in the future.

\section{FRINGE TRACKER}

Even in the optimistic case with telescopes of 8\,m diameter and $\eta = 0.35$ for a dual polarisation design, signal-to-noise at N=7.5 star on a single baseline is only $\sim$0.2 in a 100\,ms coherence time. This means that out-of-band fringe-tracking is needed, with the fringe phase corrected in post-processing, at least over a timescale of $\sim$10 seconds. This means that a companion direct-detection beam combiner is required. Once the signal-to-noise at 11.5\,$\mu$m significantly exceeds 1 (as expected in 10\,s), phase referencing at 11.5$\mu$m can be used, with closure-phase techniques. Note that water vapour seeing is not expected to be significant: a 10\,s timescale corresponds to only approximately 0.15 radians RMS phase error due to water vapour seeing at 11.5\,$\mu$m with fringe tracking at 1.6\,$\mu$m\cite{Colavita04}.
 
Our conceptual design for this companion instrument is a H-band (1.65\,$\mu$m) system with length-matched and controlled single-mode fibers for beam transport and a delay line system incorporating both a fixed and variable delay line components, and with Gaussian beam waists occurring part-way through each delay system. With a 6\,km Fresnel length for a 1.65\,$\mu$m wavelength and a $\sim$0.1\,m diameter beam, relatively small vacuum pipes ($\sim$0.2\,m diameter) can be used for the $\sim$30 fixed delay components. 

Based on simple SNR considerations and anticipated full system H-band efficiencies exceeding 1\% (including effects of fiber coupling), fringe tracking on a H=7.5 star should in principle be relatively straightforward for telescope diameters greater than 1\,m and modern, low noise detectors\cite{Finger10}. However, with the last generation of detectors, active fringe tracking has proved moderately difficult, with magnitude limits of H$\sim$7.5 on the challenging end for the VLTI with the 1.8\,m Auxiliary telescopes\cite{Sahlmann10}, but certainly anticipated as routine with adaptive optics and new detectors. For this reason, a minimum telescope diameter for 2\,m is needed for PFI.

The fringe tracker itself should almost certainly neither be a pairwise nor an all-in-one combiner.  Baselines longer than $\sim$2\,km over-resolve solar-type young stars in nearby star forming regions ($\sim$0.15\,milli-arcsec diameters), so the longest baselines would not be useful for fringe tracking. Given the likelihood of at least one telescope failure, a pairwise combiner is also likely not optimal, with an appropriate compromise being one where each telescope is combined with $\sim$4 other telescopes (generally nearest neighbours). This would still enable data from the fringe tracker to be used for short-wavelength imaging at a spatial resolution exceeding the long wavelength light.

\section{COST CONSIDERATIONS}

The largest clear identifiable cost for the array is the telescopes themselves. 30 telescopes of 4\,m diameter is a total size comparable to the Giant Magellan Telescope (GMT) in terms of collecting area. Given the scaling laws of van Belle et al.\cite{vanBelle04}, one might expect the cost of the telescope component for many 4\,m telsscopes to be smaller by a factor of $\sim$2 than GMT, especially given that the telescopes only have to be designed once and constructed many times. In any case, as long as the ratio between instrumentation and telescope costs are not so different for PFI as for extremely large telescopes, a total construction cost in the range 0.5 to 1\$B USD appears plausible. The lower limit on the cost envelope for PFI is the moving mass (structural steel), glass (area) and building (high end commercial space) costs, which is of order 0.1\$B USD - this is of course unrealistically small. 

\section{DIRECT-DETECTION COMPARISON}

For the same bandwidth and an assumed efficiency $\eta=0.35$, there is a factor of $\sim$10 improvement in signal-to-noise for direct detection interferometry over heterodyne interferometry for a single telescope\cite{Hale97}. This efficiency may be unrealistic for direct detection in particular, given typical overall efficiencies of interferometers significantly lower than this. A direct detection interferometer could be highly optimised in principle, with bare gold gratings giving 99\% reflectivity at mid-infrared wavelengths and losses due to surface roughness being negligible. 

Where an interferometer needs to image a complex scene, direct detection suffers from an additional loss where each beam has to be split many ways in pairwise combination, or where an all-in-one combiner, one has to measure signal from each pair of telescopes in the presence of photon and background noise from all other telescopes. We will assume that an all-in-one combiner is used: this additional noise factor is roughly 
$\sqrt{N_T}$, meaning that with 30 telescopes, direct detection still retains a factor of $\sim$2 advantage in principle over heterodyne interferometry. 

However, the beam-combiner could not have a very simple design, as the field of view for PFI requires a spectral resolution of at least R$\sim$400. One example design would be an integral field unit with 100 spectral resolution elements behind a 30-aperture non-redundant aperture mask. The complexity of such a system is roughly equivalent to the high resolution spectrograph needed for heterodyne interferometry. Hence, a first-order assumption is that with the exception of the additional $\sim$10--15 mirror beam train, the direct-detection and heterodyne techniques have the same efficiency. Overall, the direct detection scheme might then (optimistically) have $\sim$80\% of the dispersed heterodyne technique's efficiency.

A cost unique to the direct detection scheme is the vacuum pipes for beam transport. The maximum pipe length of 4\,km, has a corresponding Fresnel length of 0.21m at 11.5$\mu$m. Assuming no optics along the pipe, this would require a minimum pipe inner diameter of $\sim$0.4\,m. In the beam-combining laboratory, dual off-axis paraboloid mirrors specific to each telescope would need to re-image the pupil and send the beams into the vacuum delay lines. 

The delay line beam size is also required to be scaled up with cross sectional area proportional to wavelength, with a beam diameter of at least  $\sim$0.3\,m as well, perhaps requiring a 0.7\,m diameter vacuum pipe for the delay line. Assuming a cost scaling proportional to cross-sectional area, delay lines appropriate to 11.5\,$\mu$m would have $\sim$7 times the cost of delay line appropriate to H-band (1.65\,$\mu$m). 

\section{CONCLUSIONS}

We have demonstrated that heterodyne interferometry is a plausible competitor to direct detection interferometry for the Planet Formation Imager concept. Critical to the design is mixing starlight with a frequency comb laser in each telescope, dispersing the light in a high-resolution spectrograph and detecting the light mixed with each line of the frequency comb individually. Fringe tracking using direct detection interferometry is still required, but this is significantly simplified by transporting the beams from the telescopes to the delay lines by fiber. The overall cost of the project is likely comparable to the low-end of ELT budgets.

\acknowledgments     
 
M. Ireland would like to acknowledge support from the Australian Research Council under the Future Fellowships scheme (FT130100235). 


\bibliography{../mireland}   
\bibliographystyle{spiebib}   

\end{document}